\documentclass{aa}
\usepackage{natbib}
\usepackage{graphicx}
\begin{document}

\title{Spectral Behavior of Solar Oscillations Modulated \\ 
by Magnetic Variation}

\author{Heon-Young Chang}

\offprints{Heon-Young Chang}

\institute{
Korea Institute for Advanced Study \\
207-43 Cheongryangri-dong Dongdaemun-gu, Seoul 130-012, Korea\\
\email{hyc@ns.kias.re.kr}
}

\date{Received ; accepted }

\abstract{
The principal aim of observational helioseismology is to
determine mode parameters of the solar oscillations as accurately as possible. 
Yet estimates of the mode parameters are subject to many sources of
noise, including inevitable gaps in data, stochastic nature of 
excitation processes, interferences among modes. Another uncertainty
in frequency determination results from a modulation of the oscillation
frequency in itself. It is well known that variation in the mean strength 
of the solar magnetic field modulates the frequency. 
We consider effects of the solar magnetic field variation on the 
spectral behavior of the power spectrum using a simple model. 
We show that
the solar magnetic field variation can cause unwanted sidelobes
beside the main peak in the power spectrum,
and consequently show that the effect of the frequency modulation
due to the solar magnetic field variation can lead a bias in
frequency determinations.
This effect should be considered seriously particularly 
when the $ l \ne 0$ $p$-mode multiplets are analyzed to measure 
the rotational splitting since the separations in both cases could 
be comparable.
In addition to a bias in frequency estimates the line width is also likely
to be overestimated due to a line broadening, which is a measure of
the life time of the mode. 
We therefore suggest that determination of the mode 
parameters should be derived with due care considering the magnetic 
field variation when a long data set is analyzed.
The frequency modulation should be 
taken into account in the analysis to make a solid
conclusion on any subtle dependence of the mode parameters, 
such as, that on the solar cycle. We emphasize, however, that
a realistic model of the magnetic activity
variation over the entire surface of the Sun should be involved in order to
predict the effect of the solar magnetic field variation
on the spectral profiles of the solar oscillations at the quantitative level.
Finally, we conclude by pointing out that a new method is required to 
accommodate the stochastic force and the phase variation.
}

\authorrunning{Chang}
\titlerunning{Spectral Behavior of Solar Oscillations Modulated by 
Magnetic Variation}
\maketitle

\section{Introduction}

Helioseismology is a study of the solar oscillations sounding 
the internal structure of the Sun (e.g., Deubner \& Gough 1984). 
Frequencies of the solar $p$-modes provide the most 
fundamental and important information on the Sun in that one may infer 
the solar internal structure and rotation from the observed frequencies 
(Gough 1984; Christensen-Dalsgaard et al. 1985). To obtain accurate
frequencies,  it is conventional to take a Fourier power spectrum 
of time series data obtained from velocity/luminosity variation 
observations and to measure the frequencies of the peaks in the spectrum
by fitting a symmetric Lorentz profile to the observed power spectrum, 
assuming that those frequencies
are constant over the time interval encompassed by the spectrum.
This method also yields estimates of the line width and amplitude
of the solar oscillations, which include clues of their damping and 
excitation  mechanisms.

The observed power spectrum of the solar oscillations is, however, far from 
an ensemble of smooth curves. Gaps in the time series, such as, due to 
the diurnal rotation of the Earth, produce deleterious sidelobes in the 
power spectrum. The noisy power spectrum requires elaborate data reduction 
techniques to compensate  to some extent for effects of gaps in the data  
(Brown \& Christensen-Dalsgaard 1990; 
Lazrek \& Hill 1993; Chang \& Gough 1995; Fossat et al. 1999; 
Fierry Fraillon \& Appourchaux 2001). 
Even if continuous data sets without gaps exist, there is still 
an uncertainty in determining mode parameters due to the stochastic 
nature of the excitation process 
of the solar oscillations (Goldreich \& Keeley 1977; Goldreich \& Kumar 1988). 
Woodard (1984)  showed  empirically that for the case of 
a harmonic oscillator excited  by random noise the power spectrum 
will be distributed as the $\chi^{2}$ distribution with two degrees of freedom 
(for rigorous discussion see Gabriel 1994). That is, the standard  deviation
of the power at a certain frequency is equal to the mean power 
at that frequency. What it implies is that
even if the length of an uninterrupted data set becomes extended
the observed power spectrum may not converge to a smooth curve.
Instead the observed power spectrum is likely to be more spiky while
the resolution is improved. Alternatively, one may attempt to
measure the mode parameters by dividing the whole data set into several
shorter data sets and averaging  obtained spectra before the fit,
since the statistics may become more nearly Gaussian (e.g., Sorensen 1988). 

Unfortunately, however, this could not be a solution in analyzing 
the real solar power spectrum. 
The solar structure undergoes a slow change, causing variation 
in the frequency. The average of power spectra is simply an average of spectra
at different epochs with different frequencies. Therefore, strictly speaking, 
the obtained frequency from the averaged power spectrum is an average of varied 
frequencies, and the measured line width is likely to be over-estimated due to 
such a frequency variation.
Woodard \& Noyes (1985) first reported a significant decrease in frequency 
using solar intensity data from  ACRIM (Active Cavity Radiometer 
Irradiance Monitor), which were obtained from 1980 
(near solar maximum) to 1984 (near solar minimum). Since then the correlation 
between the frequency shift and the solar cycle is firmly established 
(Libbrecht\& Woodard 1990; Woodard et al. 1991; Bachmann \& Brown 1993; 
Elsworth et al. 1994; R\'egulo et al. 1994; 
Chaplin et al. 1998; Bhatnagar, Jain, \& Tripathy 1999; 
Howe, Komm, \& Hill 1999; 
Jain, Tripathy, \& Bhatnagar 2000; Chaplin et al. 2001).  
Careful interpretation of frequency dependence on the variation 
conclusively suggests that the frequency variation is mainly due to 
the perturbation of the  sound speed  near the solar surface  
(Goldreich et al. 1991; Balmforth, Gough, \& Merryfield 1996;
Dziembowski, Goode, \&  Schou 2001). 

In this paper, we demonstrate that the variation of the solar magnetic
field strength which leads to the frequency modulations of the solar
acoustic modes may cause extra noisy peaks in the observed power spectrum
in addition to the effects of stochastic excitation and a line to broaden. 
A general introduction of the frequency modulation can be found in 
radio communication engineering (e.g., Panter 1965). In helioseismology 
context the solar magnetic field variation 
and $g$-modes were considered as an agent of 
the frequency modulation (Kennedy, Jefferies, \& Hill 1993; 
Chang 1996; Lou 2001). Previous studies have been concentrated on
effects due to the existence of $g$-modes or issues on a time series analysis. 
We here consider effects of the solar magnetic field variation on the 
spectral behavior of the power spectrum using a model 
and also show that the effect of the frequency modulation
due to the solar magnetic field variation depends on the solar cycle,
as expected in the observed correlation between the frequency shift and
the solar cycle. The paper begins with a description of the 
frequency modulation and presents results of our simulations in Sect. 2.
We discuss the effects of the solar magnetic variation on the observed
spectrum and conclude in Sect. 3.

\section{Frequency-Modulated Solar Acoustic Modes}

As Lou (2001) adopted, one may derive the solar $p$-mode frequency shift 
due to the solar magnetic modulation in terms of the variational 
principle formalism (Chandrasekhar 1964). The instantaneous angular 
frequency $\omega(t)$ of a particular $p$-mode modulated 
by the solar magnetic field variation can be given as 
$\omega(t)=\omega_{0}+\epsilon f(t)$, where  $\omega_{0}$ is the 
unmodulated angular frequency\footnote{an angular frequency $\omega_0$ 
is 2 $\pi$ times a cyclic frequency $\nu_0$}, $\epsilon$ is the strength
of the deviation which is assumed to be related to 
mode properties, such as, inertia, and $f(t)$ is a function which is assumed to  
vary with the  solar magnetic field strength. We note 
that  $\epsilon$ is so small that the modulus of $\omega_{0}$ is much 
greater than that of $\epsilon f(t)$, and $f(t)$ is a slowly varying 
function compared to the  $p$-mode and same for all the solar $p$-modes 
since all the solar 
oscillations contributing to the signal are experiencing the same 
perturbation agent to the acoustic cavity. We used solar magnetic field 
data\footnote{ ${\rm ftp://argo.tuc.noao.edu/kpvt/daily/stats/mag.dat}$} 
available at the National Solar Observatory at Kitt Peak to generate 
$f(t)$, converting the observed magnetic field strength to the frequency 
variation with the linear relation obtained by Woodard et al (1991).
Though the apparent periodicity is most likely related to
presence of active regions that corotate with the Sun, we 
presume that this feature simulate
intrinsic periodic variations of solar magnetic field
strengths that may lead to
frequency modulations of solar p-modes. It should be pointed out, however,
that one should employ a more detailed model for the distribution of 
magnetic field in the solar interior to estimate actual
biases of frequency and line width determinations.
The relation between the magnetic field strength and the frequency
shift may not be an optimal function (Bhatnagar et al. 1999), 
but any subtle discrepancy should not be
a serious problem in this demonstration. Since Woodard et al (1991) normalized 
the relation to $l=0$ mode of $ \approx 3 $ mHz, in our simulation
$\epsilon$ may vary a factor of two depending on $\omega_0$ we take.
The strength of the deviation  $\epsilon$ is roughly an increasing
function with the frequency.
The  simulated signal of the frequency-modulated $p$-mode can be
written as
\begin{equation}
I_{\rm FM}(t)= 
A (t)\cos (\omega_{0} t + \epsilon \int^t_{0} f(\tau ) d\tau + \psi_{0}),
\end{equation} 
where $A(t)$ is an amplitude which we consider to be  determined 
such that the energy distribution follows the Boltzmann distribution, and 
$\psi_{0}$ is a phase when $t=0$. Without loss of 
generality the phase constant can arbitrarily be set equal to 
zero and $A(t)$ constant $A_0$.

%Figure 1
\begin{figure}
\resizebox{\hsize}{!}{\includegraphics{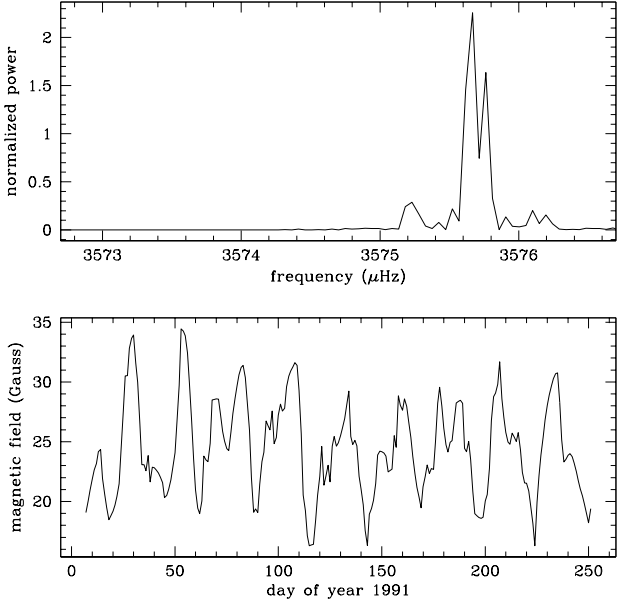}}
\caption{
Simulated power spectrum of the frequency-modulated mode is shown
in the upper panel.
The spectral line is significantly broadened and apparent 
sidelobes are spaced at $\sim 0.4~ \mu$Hz.  
Note that power is arbitrarily normalized and noise-free.
 The frequency modulation function $f(t)$ is
generated from the magnetic field strength observed in 1991 
(near solar maximum) which is shown in the lower panel. 
}
\end{figure}

Since an arbitrary continuous function $f(t)$ can be expressed 
in terms of harmonics, according to Fourier's theorem, we begin with 
a single-tone sinusoid as an illustration, such that $f(t) = \cos \omega_{m} t$. 
Then the frequency-modulated function is given by 
\begin{equation}
I_{\rm FM}(t)  = A_{0} \cos (\omega_{0} t +  \beta \sin \omega_{m} t),
\end{equation} 
where $\beta$ is $\epsilon/\omega_{m}$ so that the maximum phase deviation 
is inversely proportional to the frequency of the modulating function. 
The representation for $I_{\rm FM}(t)$ can be expanded  due to Jacobi's 
expansion in a series of Bessel coefficients using the Fourier series 
expansion (Watson 1922). Using  a property of Bessel functions we have
\begin{equation}
I_{\rm FM}(t)= 
A_{0} \sum^{\infty}_{n=-\infty}J_{n}(\beta) \cos (\omega_{0} t + 
n \omega_{m} t ),
\end{equation}
where $J_{n}(\beta)$ is the Bessel function of the first kind
with  $J_{n}(\beta)= (-1)^n J_{-n}(\beta)$. 
Thus we have a time function consisting of a main unmodulated 
function and an infinite number of sidebands, 
whose relative amplitudes are proportional to $J_{n}(\beta)$ 
spaced at frequencies $\pm n \omega_{m}$. 
As $\beta \rightarrow \infty$, the number of sidebands increases 
and the spectral components become more and more confined between 
$\omega_{0} \pm \epsilon$. And the maximum height is also reduced because 
energy tends to be distributed among many small peaks. Note that $\beta$ 
is proportional to the sensitivity of the mode on the solar magnetic 
variation. In reality, the modulating function is multitones in
nature. The signal suffers more interference, causing more sidebands.
Obviously  the power spectrum of a modulated function depends on 
a shape of the modulation function. For instance, in the case 
where the frequency modulation is a square wave function, when 
the amplitude of the square wave $\bigtriangleup \omega$ is much 
larger than the fundamental frequency of the square wave the power 
spectrum shows two distinctive peaks in the vicinity of 
$\omega_{0} \pm \bigtriangleup \omega$.

%Figure 2
\begin{figure}
\resizebox{\hsize}{!}{\includegraphics{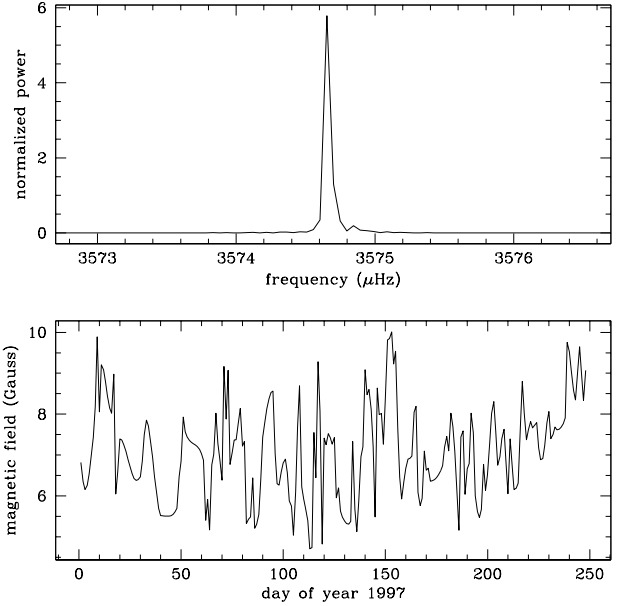}}
\caption{
Similar plots with Figure 1, except that the  frequency modulation function
is derived from the data obtained in 1997 (near solar minimum).
Sidelobes are less significant than those in Figure 1, yet the spectral
line is still broadened.
}
\end{figure}

In Figures 1 and 2 we show the Fourier power spectra of simulated data 
generated by Equation (1). In our simulations we presume $l=0$ mode 
with its cyclic frequency 3574.7 $\mu$Hz and set $\epsilon=2$ as the 
effect of the frequency-modulation  should be more serious in higher 
frequency modes. The  duration of the observation corresponds to 8 months. 
As mentioned earlier, we take the observed magnetic field strength and 
translate it into $f(t)$ according to the linear relation given by 
Woodard et al (1991).  We show two specific cases of
different epochs, that is, year 1991 (near solar maximum) and 
1997 (near solar minimum) to demonstrate its dependence on the solar cycle.
The observed magnetic field strengths we adopt have been shown in 
the lower panels separately in Figures 1 and 2. 
Besides the obvious frequency shifts, which can be defined 
such that $T^{-1} \int^T_0 \omega(t) dt -\omega_0$ where
$T$ is the interval of observation, unexpected sidelobes appear (Figure 1).
In the case of year 1991, the solar magnetic modulation is almost
a single-tone sinusoid with $\nu_m \approx 0.4 ~\mu$Hz, which is 
a similar value of the amplitude of $\epsilon f(t)$ in Equation (2). 
In other words, the peak
phase deviation $\beta$ is $\sim 1$. This fact results in sidelobes
located at $\sim  0.4 ~\mu$Hz apart from the main peak.
The effect of the frequency modulation is likely more conspicuous in the
case of the solar maximum. 
It should be noted, as we mentioned above, that these specific sidelobes 
at $\sim 0.4 ~\mu$Hz may not be manifest in the real solar 
power spectrum since the magnetic variation used here with periods 
of $\sim 30$ days results from the apparent periodicity due to
the presence of active regions corotating with the Sun. 
In Figure 2, the only possibly detectable effect of 
the frequency modulation is a broadening of the main peak.  
In this particular case, the spectral line is more broadened by a factor of two
than that in the case where there is no frequency-modulation.
One important thing to bear in mind is, therefore, that according to the 
simulated power spectrum the spectral line width is
likely to be overestimated in any case if the effect of the frequency modulation
is ignored. Moreover, when the mode is frequency-modulated the height of
the central peak is also likely reduced since the energy of the mode 
tends to be distributed among many small sidelobes as shown above. We note that
a decrease in the strength of the modes from solar minimum to maximum is  
reported by the BiSON group, where the effect of the frequency modulation
has been ignored in the analysis (Elsworth et al. 1993). 
The frequency modulation should be taken into account to make a solid
conclusion on this matter. 
Otherwise, estimates of the life time and the strength of 
the solar oscillations seems likely to be biased.

%Figure 3
\begin{figure}
\resizebox{\hsize}{!}{\includegraphics{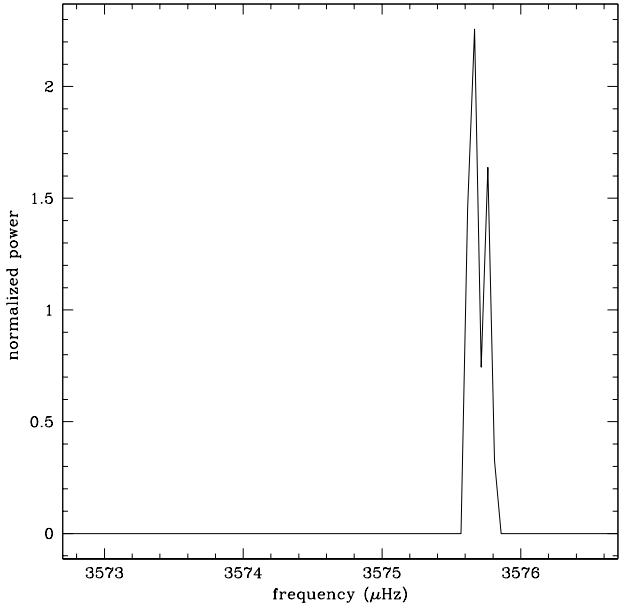}}
\caption{
Similar power spectrum shown in Figure 1, except that all other
noisy structures further than $\pm~ 0.2 ~\mu$Hz from the main peak are 
set to zero.
}
\end{figure}
 
\section{Discussion and Conclusion}

Magnetic activity changes the outer layers of the solar envelope, 
modifying the resonant properties of the $p$-mode cavity and modulating 
the oscillation frequencies. By simply measuring the positions of peaks 
in power spectra, one is restricted by the uncertainty in frequency 
determination due to the phase wandering induced by the changing cavity.
We demonstrate that the frequency modulation by the solar magnetic
field variation may affect estimates of the mode parameters, particularly
during a period near the solar maximum. This effect should be seriously
taken into consideration when the  $ l \ne 0$ $p$-mode multiplets are 
analyzed to measure the rotational splitting. 
As shown above the sidelobes due to the solar magnetic variation
may be displaced next to the main peak at similar amounts of the rotationally
splitting separation if the magnetic field varies 
with a roughly 30-day timescale. 
Although many attempts to correlate the solar 
rotation curve with the solar cycle have been made 
(Jim\'enez et al. 1994; Chaplin et al. 1996; Antia \& Basu 2000), 
we suggest that conclusions of the dependence of the solar
rotation curve on the solar cycle should be derived with due care
since the mode estimatation could be biased due to the frequency modulation
as well as the mode interference (Appourchaux et al. 2000).
In addition to a bias in frequency estimates, the line width is also likely
to be overestimated. In Figure 3, we show the power spectrum shown
in Figure 1, in which  all noisy structures further than $\pm ~0.2 ~\mu$Hz 
from the main peak are set to zero, in order to emphasize the line broadening
and to suppress the possible rotational effect.

We conclude by pointing out that
there was an attempt for developing a method to measure the frequency
by fitting the temporal signal to a physical model deduced by 
the superposition of many modes, 
rather than using its power spectrum (Chang 1996; Chang \& Gough 1995). 
This technique uses the fact that the unknown components to the wandering 
of amplitudes and phases of constituent oscillators arising from 
the temporal variation of the cavity in which they are confined are 
related to each other in a known way. Provided that a sophisticated algorithm
is implemented to accommodate the stochastic force and the 
surface variation causing the phase modulation, such an idea may yield
more accurate estimates of  the frequency and the life time of the solar 
short period oscillations.

\acknowledgements{
We would like to thank Douglas Gough and Takashi Sekii for useful discussions
while the idea is conceived, and the anonymous referee for critical
comments which improve the original manuscript.
NSO/Kitt Peak magnetic data used here are produced cooperatively by
NSF/NOAO, NASA/GSFC and NOAA/SEL.}

\end{document}